\documentclass[conference]{IEEEtran}
\IEEEoverridecommandlockouts
\usepackage{cite}
\usepackage{amsmath,amssymb,amsfonts}
\usepackage{algorithmic}
\usepackage{graphicx}
\usepackage{textcomp}
\usepackage{xcolor}
 \usepackage{float}
\usepackage{amsmath}
\usepackage{latexsym}
\usepackage{amsmath}
\usepackage{amsfonts}
\usepackage{amssymb}
\usepackage{mathdots}
\usepackage{comment}
\usepackage{float}
\usepackage[ruled,vlined]{algorithm2e}
\usepackage{algorithmic}
\usepackage{caption}
\usepackage{adjustbox}
\usepackage{etoolbox}
\usepackage{setspace}

\usepackage{multirow}
\def\BibTeX{{\rm B\kern-.05em{\sc i\kern-.025em b}\kern-.08em
    T\kern-.1667em\lower.7ex\hbox{E}\kern-.125emX}}
    
\begin{document}

\title{Fast and Scalable Complex Network Descriptor Using PageRank and Persistent Homology}

\author{\IEEEauthorblockN{1\textsuperscript{st} Mustafa Hajij}
\IEEEauthorblockA{\textit{Department of Mathematics and Computer Science} \\
\textit{}\\
Santa Clara, California\\mhajij@scu.edu}
\and
\IEEEauthorblockN{2\textsuperscript{nd} Elizabeth Munch }
\IEEEauthorblockA{\textit{ Department of Computational Mathematics, Science, and Engineering} \\
\textit{Michigan State University}\\
Lansing, Michigan \\
muncheli@msu.edu}
\and
\IEEEauthorblockN{3\textsuperscript{rd} Paul Rosen}
\IEEEauthorblockA{\textit{Department of Computer Science and Engineering} \\
\textit{University of South Florida}\\
Tampa, Florida \\
prosen@usf.edu}
}

\maketitle

\begin{abstract}
The PageRank of a graph is a scalar function defined on the node set of the graph which encodes nodes centrality information of the graph. 
%In this work, we utilize the PageRank function on the lower-star filtration of the graph as input to persistent homology to study the problem of graph similarity. 
 In this article we use the PageRank function along with persistent homology to obtain a scalable graph descriptor and utilize it to compare the similarities between graphs. For a given graph $G(V,E)$, our descriptor can be computed in $O(|E|\alpha(|V|))$, where $\alpha$ is the inverse Ackermann function which makes it scalable and computable on massive graphs.
 We show the effectiveness of our method by utilizing it on multiple shape mesh datasets.
\end{abstract}


\section{Introduction}

 The problem of studying similarity between graphs has attracted much attention recently in the pattern recognition and machine learning communities. 
 One of the main challenges is to construct an effective similarity measure between graphs that takes into account the complexity of the underlying structure while still being computed efficiently.  

%  One of the main challenges in graph similarity is to construct construct an effective similarity measure that takes into account the complexity of the underlying structure and can be computed efficiently.  

In this work, we utilize the \textit{PageRank} vector \cite{BrinPage1998} in conjunction with a tool available in \textit{persistent homology} \cite{edelsbrunner2000topological} to define a graph descriptor. 
More specifically, we view the PageRank as a continuous scalar function \cite{Pretto2008} defined on the vertices of the graph and utilize this scalar function to induce a \textit{filtration} as defined traditionally in the context of persistent homology. 
We show that the \textit{persistence diagram} induced by this filtration can be utilized for graph similarity.

Persistent homology provides a robust set of tools for the theoretical and practical capacity to understand the \textit{shape of data} \cite{carlsson2009topology} in any number of dimensions and on multiple scales, placing the concept of shape, as applied to data analysis, on a solid mathematical foundation. 
On the other hand, the PageRank function of a graph stores information regarding the centrality information of the underlying nodes. 
The filtration induced by the PageRank provides a method to decode the information encoded in this scalar function and stores it in the persistence diagram. 
The latter, when combined with bottleneck distance, can then be used for the graph similarity task. 

Utilizing the PageRank vector has two main advantages. First, PageRank was originally designed to compute efficiently on very large graphs. 
The efficiency of the PageRank vector has been studied extensively \cite{haveliwala1999efficient}. The PageRank vector has found many applications, including graph partition \cite{andersen2006local}, image search \cite{jing2008pagerank}, and citation analysis \cite{ma2008bringing}, among others.
Second, as we will show here, as a function defined on the nodes of the graph the PageRank vector stores rich structural information about the underlying graph that can be utilized to to detect the similarity between different graphs effectively.

Graph similarity lies within the realm of pattern recognition and machine learning \cite{rehman2012graph}. Persistent homology provides unique information about the graphs, discover uncovering insights, and determines which predictors are more related to the outcome. Persistent Homology-based methods have shown excellent performance in several applications including pattern recognition on graphs~\cite{ CarstensHoradam2013, ELuYao2012, suh2019persistent, PetriScolamieroDonato2013, PetriScolamieroDonato2013b}, time-varying data~\cite{edelsbrunner2004time, hajij2018visual}, and images \cite{clough2019topological,garside2019topological,robles2017shape}, among others.

\section{Background}
In this section, we give a brief review of persistent homology and the PageRank vector. While the work here is concerned with graphs, we choose here to introduce persistent homology for  simplicial complexes since our work can be generalized easily to more general domains. We assume the reader is familiar with the basics of simplicial homology.

\subsection{Persistent Homology}

% https://arxiv.org/pdf/1810.04807.pdf
% http://web.cse.ohio-state.edu/~dey.8/paper/DiscreteMorse/MorseGIS.pdf

%\textcolor{blue}{@ optional task : try to add the definition of a simplical complex 
%https://en.wikipedia.org/wiki/Simplicial_complex.}

Let $K$ be a simplicial complex. We will denote the vertices of $K$ by $V(K)$. Let $S$ be an ordered sequence $\sigma_1,\cdots,\sigma_n $ of all simplices in $K$, such that for simplex $\sigma \in K $ every face of $\sigma$ appear before it $\sigma$ in $S$. Then $S$ induces a nested sequence of subcomplexes called a \textit{ filtration}:
$\phi=K_0  \subset K_1 \subset ... \subset K_n = K$.
 A $d$-homology class $\alpha \in H_d(K_i)$ is said to be \textit{born} at the time $i$ if it appears for the first time as a homology class in $H_d(K_i)$. A class $\alpha$ \textit{dies} at time $j$ if it is trivial $H_d(K_j)$ but not trivial in $H_d(K_{j-1})$. The \textit{persistence} of $\alpha$ is defined to be $j-i$.  Persistent homology captures the birth and death events in a given filtration and summarizes them in a multi-set structure called the \textit{persistence diagram} $P^d(\phi)$. Specifically, the persistence diagram of the a filtration $\phi$ is a collection of pairs $(i,j)$ in the plane where each $(i,j)$ indicates a $d$-homology class that is created at time $i$ in the filtration $\phi$ and killed entering time $j$.  

 Persistent homology can be defined given any filtration. For the purposes of this work, the input is a piecewise linear function $f:|K|\longrightarrow \mathbb{R}$ defined on the vertices of complex $K$.
 \begin{figure*}[!t]
\begin{center}
\includegraphics[width=0.99\linewidth]{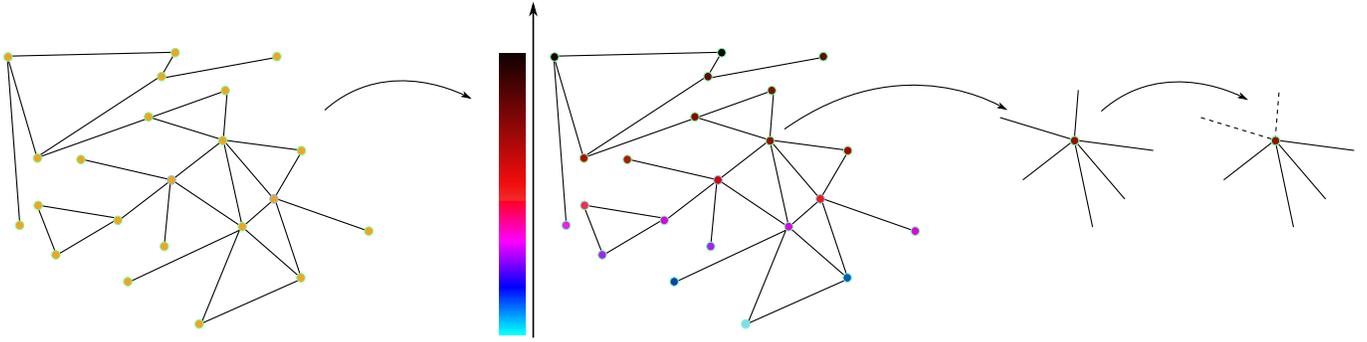}
\end{center}
\vspace{-2mm}
\caption{Left : a graph with a scalar function defined on its nodes. Middle the star  of the node $v$. Right: the lower-star of a vertex $v$.}
\label{fig:filtration_example}
\end{figure*}
 Furthermore, we assume the function $f$ has different values on different nodes of $K$. 
 Any such a function induces the \textit{lower-star} filtration as follows.

%Let $v\in V(K)$ be a vertex of $K$. 
%The \textit{star} of $v$, denoted as $St(v)$, is the set of all simplices in $K$ that contain $v$ as a vertex. 
%When we are given a piece-wise linear function $f$ defined on $K$, we can also define the lower star of $v$. 
%Namely, the lower star of a vertex $v \in V(K)$ as $LowSt(v)=\{w \in St(v)| f(w)\leq f(v)\}$. 

Let $V=\{v_1,\cdots,v_n\}$ be the set of vertices of $K$ sorted in non-decreasing order of their $f$-values, and let $K_i:= \{\sigma \in K | \max_{v \in \sigma}f(v)\leq f(v_i)   \} $. 
The lower-star filtration is defined as: 

\begin{equation}
\label{filter2}
    \mathcal{F}_f(K):  \phi=K_0  \subset K_1 \subset ... \subset K_n = K.
\end{equation} 
The lower-star filtration reflects the topology of the function $f$ in the sense that the persistence homology induced by the filtration \ref{filter2} is identical to the persistent homology of the sublevel sets of the function $f$. 
We denote by $P_f(K)$ to the persistence diagram induced by the lower-star filtration $ \mathcal{F}_f(K)$. See Figure \ref{fig:filtration_example}.

Furthermore, we will denote by $P^k_f(K)$ to the $k^{th}$ persistence diagram induced by the lower-star filtration $ \mathcal{F}_f(K)$. 
In this work,  we will only consider the 0-dimensional persistence diagram.

%Start by extending $f$ to a function $\Bar{f}: K \longrightarrow \mathbb{R}$ such that each simplex in $K$ is assigned the maximum function value of the vertices it contains.

\begin{figure*}[!t]
\begin{center}
\includegraphics[width=.82\linewidth]{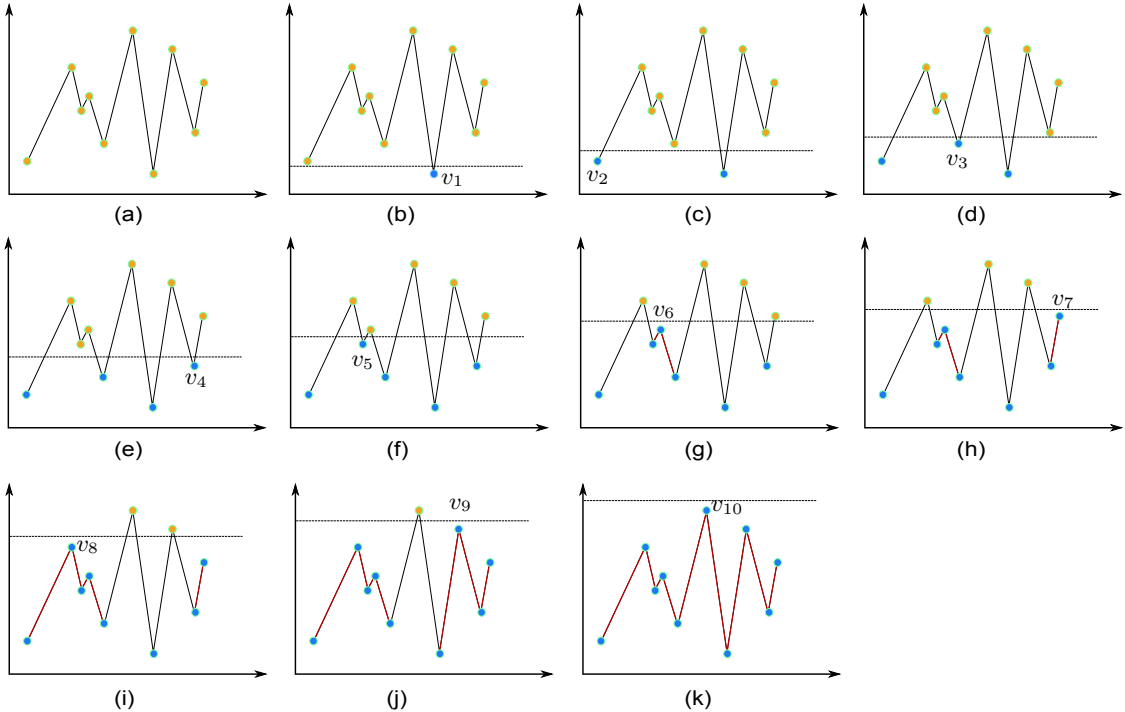}
\small{
\put(-258,199){$v_1$}
\put(-203,200){$v_2$}
\put(-67,206){$v_3$}
\put(-355,123){$v_4$}
\put(-292,130){$v_5$}
\put(-178,149){$v_6$}
\put(-27,153){$v_7$}
\put(-396,60){$v_8$}
\put(-255,76){$v_9$}
\put(-156,75){$v_{10}$}
}
\end{center}
\vspace{-2mm}
\caption{An example illustrating the computation of the persistence diagram on a scalar function defined on 1-d simplicial complex $K$. We assume that the we have a scalar function $f:V(K)\longrightarrow \mathbb{R}$ defined on the vertex set $V(K)$ of $K$. We order the nodes and the edges in the function 
using their $f$ values and process them with respect to this order. The values of the function $f$ hence induces a lower star filtration where at every stage in this filtration we introduce a vertex along with the edges that are connected to it and have lower $f$-values, if it has any. }
\label{fig:lowerstar}
\end{figure*}

\subsection{Computing the $0$-persistence diagram the
 of a lower-star filtration }
 For completeness of our treatment we give a brief description for computing the 0-persistence diagram the PageRank defined on the nodes of on a graph $G$. The computation of the zero persistent diagram $P^0_f(G)$ can actually be done  using union-find data structure. We give the details next.
 If $e=(u,v)$ is an edge of the graph $G$ then we will extend the PageRank vector to $e$ by defining $PR(e):=max(PR(u),PR(v))$.

Let $V=\{v_1,\cdots,v_m\}$ be the node set of $G$.  
 Let $E=\{e_1,\cdots,e_n\}$ be its edge set ordered with respect to their $PR$-values. The steps of the the algorithm to compute the zero PD associated with the PageRank is given as follows.
   
\LinesNumbered 

\SetKwProg{Fn}{Function}{}{}
\begin{algorithm}
\Fn{computePageRankPD($G,PR:G\longrightarrow \mathbb{R}$)}{

    $bars=[\hspace{2pt} ] $\\
    $U =\emptyset $\\
    \ForEach{Node $i$ in $V(G)$}{
            $U.make(i)$\;
        }    
    Sort the edge of the graph $G$ in ascending order using the their $PR$-values.\\
    \ForEach{Edge $e=(u,v)$ in $E(G)$}{
        
        $c \gets U.get(u)$\\
        $d \gets U.get(v)$\\
        \If{$c \neq d$}{
            $U.merge(c,d)$\\
            $bars.append( (max(PR(c),PR(d)),PR(e)) )$
        }
    }
    \Return $bars$
    }
    \textbf{End Function}

    \caption{Computing the Persistence Diagram induced by the PageRank}
     \label{alg1}
\end{algorithm}

The first step in the algorithm creates a connected component $C_i$ for each node $v_i$ in the graph $G$. Here we assume that the connected components are created using the \emph{disjoint set} data structure. 

The second step of the algorithm looks at the edges of $G$ in the ascending order with respect to their $PR$-values. For each $e=(u,v)$, we check if the nodes $u$ and $v$ of $e$ belong to two different sets. If this is the case, then we merge the two connected components containing $u$ and $v$. Furthermore, we append to the list of $bars$ the pair $(max(PR(c),PR(d)),PR(e))$ where $c$ and $d$ are the roots of the trees that contain the nodes $u$ and $v$ respectively in the disjoint set data structure. The algorithm return the list $bars$ representing the birth and death of $0$-features of the graph $G$ with respect to PageRank functional values.

The $merge$ operation in line $12$ in Algorithm \ref{alg1} assumes the following merge order on the sub-trees in disjoint set data structure. The tree with root $c$ is merged with the tree with root $d$ according to the $PR$ values of $c$ and $d$. Namely, if $PR(c)>PR(d)$ then we set $d$ to be the parent of $c$. Otherwise $c$ to be the parent of $d$.  

An illustrative example of running this algorithm on a 1-d function is given in Figure \ref{fig:lowerstar}.

\section{Computing the distance between the Persistence diagrams}
Given two persistence diagrams, we measure the distance between them using the bottleneck distance. 
Namely, given two persistence diagrams $X$ and $Y$, let $\eta$ be a bijection between points in the diagrams. 
The bottleneck distance is defined as, 
$$
W_{\infty}(X,Y) = \displaystyle \inf_{\eta: X \rightarrow Y} \sup_{x \in X} \left\lVert x-\eta(x) \right\rVert_\infty.
$$ 
For technical reasons we usually add to the persistence diagram infinitely many points on the diagonal and each one of these points with is counted with infinite multiplicity.   In our study we utilize the bottleneck distance to quantify the difference between two PR descriptors. Other distances can also be employed such as the Wasserstein distance.  
%For technichical reason we usually add To make persistence diagrams stable, each
%point (x, x) on the diagonal is counted with infinite
%multiplicity.

%{\color{red}pr: don't know if you care to mention it, but bottleneck distance requires infinite multiplicity across the diagonal. }

%Given two graphs $G_1$ and $G_2$, we can use the PageRank vector to compute a the persistence diagrams induced by  

\subsection{PageRank}

%By definition the persistence diagram is a function of underlying filtration. This filtration is typically commonly induced by computing the metric matrix of the underlying space.

 This work utilizes the lower-star filtration induced by the PageRank function \cite{BrinPage1998}; more specifically, we consider a version applicable to undirected graphs~\cite{Grolmusz2012}. 
The PageRank function $PR:V \to \mathbb{R}$ is defined for every vertex $v \in V$ by  
\begin{equation}
\label{eq:pagerank}
%\centering
PR(v)=\frac{(1-d)}{|V|}+d \sum_{ u\in  N(v)} \frac{PR(u)}{ |N(u)|}, 
\end{equation}

\begin{figure}[h]
	\centering
	\includegraphics[width=0.85\linewidth]{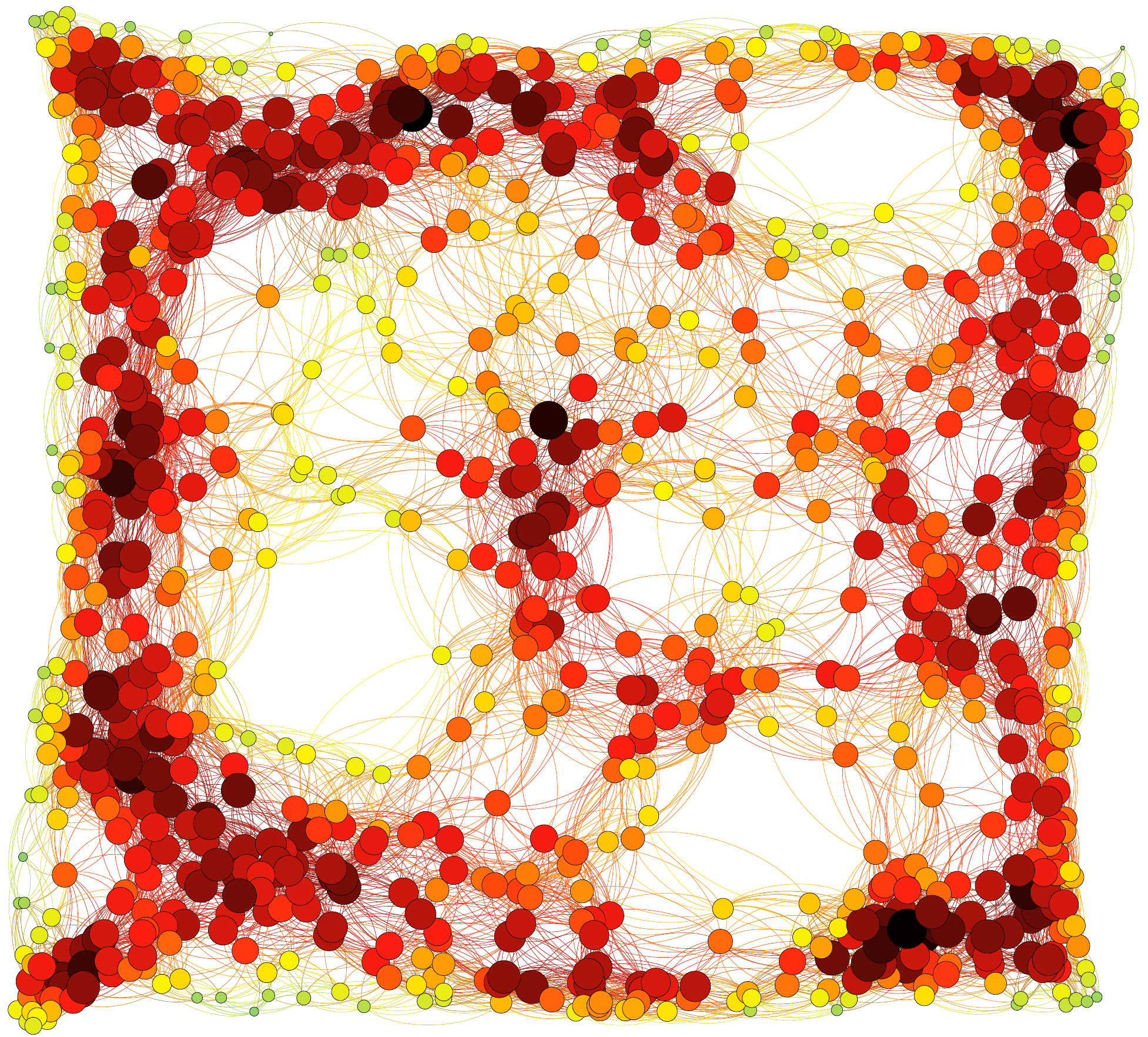}
	\caption{Example of the PageRank vector computed on a geometric graph. Higher PageRank values indicate higher node centrality. In this figure the PageRank values are indicated by the size of the nodes as well as the the color of the nodes color (nodes with higher PR values have darker colors).}
	\label{fig:pagerank}
\end{figure}

where $N(v)$ is the set of neighbors of $v$; $0<d<1$ is the \emph{damping factor}, typically set at $0.85$. 
Equation~(\ref{eq:pagerank}) can solved efficiently by the power method~\cite{HoffmanFrankel2018}.   See also \cite{sarma2013fast} for a $\mathcal{O}( \sqrt{\log(n)}/\epsilon)$ distributed algorithm where $n$ is the number of nodes in the graph and $\epsilon$ is fixed constant.

A high PageRank score at $v$ typically means that $v$ is connected to many nodes, which also have high PageRank scores. 
For our purpose, it is important to notice that the PageRank is a continuous function  ~\cite{Pretto2008}. 
For example, Figure \ref{fig:pagerank} illustrates the continuity of the function on the nodes of the graph on a random geometric graph.

%For our experimentation, we utilize the PageRank implementation in NetworkX~\cite{HagbergSchultSwart2008}. 

\section{Running Time}

The proposed descriptor can be computed in almost linear time. Once the graph data is loaded, the 0-dimensional persistence diagram can be computed  using disjoint sets which take
$O(|E|\alpha(|V|))$, where $\alpha$ is the inverse Ackermann function \cite{cormen2009introduction}, an
extremely slow growing function. The PageRank can be computed in sub-linear time. For instance see \cite{sarma2013fast} for a $\mathcal{O}( \sqrt{\log(n)}/\epsilon)$ distributed algorithm where $n$ is the number of nodes in the graph and $\epsilon$ is fixed constant. 
 
\section{Results}

%To validate the method proposed here, we test it using three publicly available data. The first 
To validate the method proposed, we run some experiments on three publicly available datasets. 
We use mesh datasets to make a visual comparison between similar graphs easier.

In our experiments, we compute the persistence diagram of each mesh obtained from the lower-star filtration induced by the PageRank vector defined on that mesh. 
The pairwise bottleneck distance is then computed between every pair of persistence diagrams.  
Finally, the resulting discrete metric space is visualized using a 2d t-SNE projection \cite{vanDerMaaten2008}.

%To validate the effectiveness of the topological descriptor
%proposed here, we test it using two publicly available data sets \cite{rodola2014dense}.

 The first dataset \cite{sumner2004deformation} consists of 60 meshes that are divided into
6 categories: cat, elephant, face, head, horse, and lion. 
Each category contains ten triangulated meshes. 
The result is reported in Figure \ref{fig:tnse projection} left handside.
\begin{comment}

\begin{figure}[H]
	\centering
	\includegraphics[width=0.99\linewidth]{combined_60_dataset_final.png}
	\caption{The application of our method to a data set consists of $60$ triangulated meshes divided into $6$ categories.}
	\label{fig:60datset}
\end{figure}  
\end{comment}

The second dataset \cite{rodola2014dense} consists of 30 meshes that are divided into 2 categories: kid A and kid B. The result is reported in Figure \ref{fig:tnse projection} right-handside.

The third dataset  \cite{bronstein2008numerical} contains a total of 80 objects, including 11 cats, 9 dogs, 3 wolves, 8 horses, 6 centaurs, 4gorillas, 12 female figures The vertex count for each object in this data is about $50K$. 

In all of our three example datasets, one can clearly observe the effectiveness of the proposed descriptor at capture the geometry of the underlying meshes. In particular, one can easily see that the meshes within the same category are clustered together. We also notice that meshes with similar topology tend to be closer than those with different topology. Observe for instance the clusters of horses and cats in Figure \ref{fig:tnse projection}.

\begin{comment}

\begin{figure}[H]
	\centering
	\includegraphics[width=0.99\linewidth]{combined_kid_final.png}
	\caption{ The application of our method to kids dataset, which consists of $30$ meshes: $15$ meshes of kid $A$ and $15$ meshes of kid $B$.}
	\label{fig:kidsdatset}
\end{figure}  

\end{comment}

\begin{figure*}[!t]
	\centering
	\includegraphics[height=95pt]{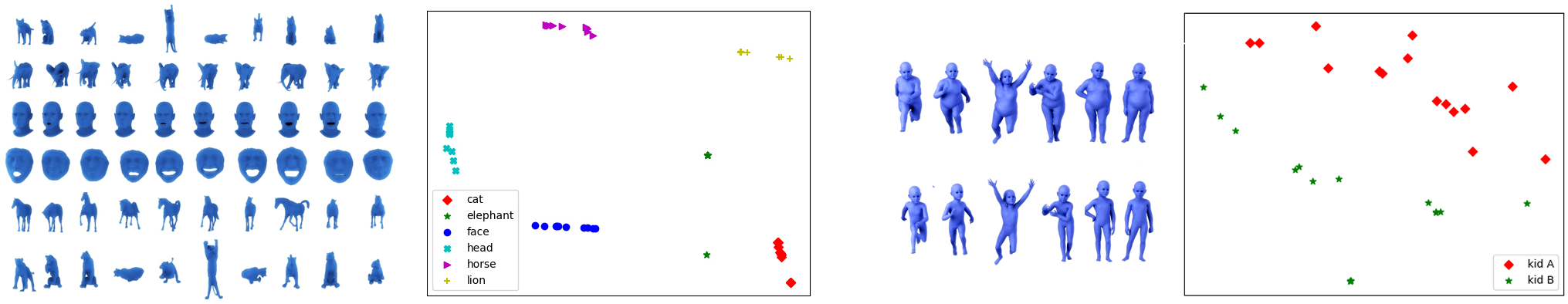}
	\caption{In both left and right Figure we compute the PageRank's vector for each mesh in a data set is computed. We then utilize this function to compute $0$-persistence diagram associated with the lower-star filtration of PageRank. Then we compute the pairwise bottleneck distance between every pair of of that dataset. The final distance matrix is then visualized using a 2d t-SNE projection.
	In the left figure, we show the application of our method to a data set consists of $60$ triangulated meshes divided into $6$ categories \cite{sumner2004deformation}. On the other hand the right figure shows the application of this method to kids dataset \cite{rodola2014dense} which consists of $30$ meshes, $15$ meshes of kid $A$ and $15$ meshes of kid $B$.}
	\label{fig:tnse projection}
\end{figure*} 

\begin{figure*}[!t]
	\centering
	\includegraphics[height=190pt]{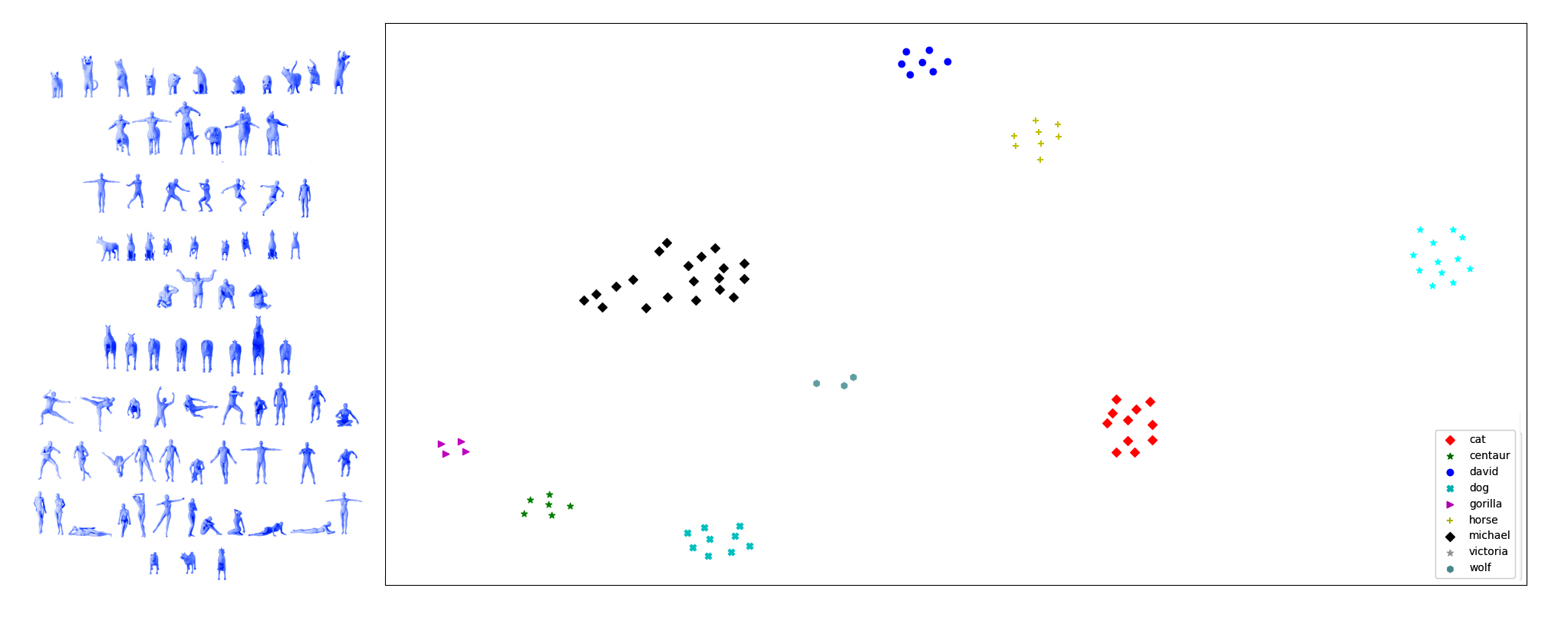}
	\caption{	
On the left the dataset  \cite{bronstein2008numerical} which consists a total of 80 objects, including 11 cats, 9 dogs, 3 wolves, 8 horses, 6 centaurs, 4 gorillas, 12 female figures. The vertex count for this dataset is about 50,000. On the right the t-SNE projection obtained from the distance matrix of the pairwise bottleneck distance between the persistence diagrams associated with the lower-star filtration of the PageRank vectors.}
	\label{fig:tnse projection}
\end{figure*}

\section{Conclusion}

In this work, we have illustrated how the PageRank can be utilized in conjunction with persistent homology to study graph similarity and demonstrated our results on small datasets. 
In future work, we are planning to conduct a more thorough analysis with larger datasets. 
Moreover, the PageRank is typically defined on directed graphs. This feature of the PageRank vector can be utilized to induce a filtration that is sensitive to the directionality of the edges a directed graph. 
We are planning to investigate this direction in the future.

\section{Acknowledgment }
This work was supported in part by a grant from the National Science Foundation (IIS-1845204).
\bibliographystyle{plain}
\bibliography{refs}
\end{document}